\documentclass[%
 reprint,
superscriptaddress,
nofootinbib,
nobibnotes,
notitlepage,
 amsmath,amssymb,
aps,
onecolumn,
prl,
]{revtex4-1}

\usepackage{graphicx}% Include figure files
\usepackage{dcolumn}% Align table columns on decimal point
\usepackage{bm}% bold math
\usepackage[colorlinks]{hyperref}% add hypertext capabilities
\def\mnras{\rm{MNRAS}}

\def\nat{\rm{Nature}}
\def\aap{\rm{A\&A}}

\def\apj{\rm{ApJ}}                 % Astrophysical Journal
\def\jcap{\rm{JCAP}}

\usepackage{color}
\usepackage{xspace}
\usepackage[caption=false]{subfig}
\def\beq{\begin{eqnarray}}
\def\eeq{\end{eqnarray}}

\usepackage[normalem]{ulem}
\usepackage{physics}
\usepackage{ifthen}
\usepackage{dcolumn}

\newcommand{\Mpch}{h^{-1}\mathrm{Mpc}}
\newcommand{\hMpc}{h\,\mathrm{Mpc}^{-1}}
\newcommand{\hun}{\,\mathrm{km}\,\mathrm{s}^{-1}\mathrm{Mpc}^{-1}}

\newcommand{\dve}[4][1]{
\ifthenelse{\equal{#1}{1}}{\left.\dv{#2}{#3}\right\vert_{#4}}{\left.\dv[#1]{#2}{#3}\right\vert_{#4}}}

\newcommand{\pdve}[4][1]{
	\ifthenelse{\equal{#1}{1}}{\left.\pdv{#2}{#3}\right\vert_{#4}}{\left.\pdv[#1]{#2}{#3}\right\vert_{#4}}}

\definecolor{darkgreen}{RGB}{0,120,0}
\definecolor{brown}{RGB}{120,60,0}

\newcommand{\resub}[1]{#1}
\usepackage[T1]{fontenc}

\usepackage[utf8]{inputenc}
\usepackage{newunicodechar,graphicx}
\DeclareRobustCommand{\okina}{%
  \raisebox{\dimexpr\fontcharht\font`A-\height}{%
    \scalebox{0.8}{`}%
  }%
}
\newunicodechar{ʻ}{\okina}
\newcommand*\hawaii{Hawai\okina{}i}

\begin{document}

%\preprint{APS/123-QED}

\title{\Large Determining the Hubble Constant without the Sound Horizon:\\\
\large A $3.6\%$ Constraint on $H_0$ from Galaxy Surveys, CMB Lensing and Supernovae}% Force line breaks with \\
\author{Oliver H.\,E. Philcox}\email{ohep2@cantab.ac.uk}
\affiliation{Center for Theoretical Physics, Department of Physics,
Columbia University, New York, NY 10027, USA}
\affiliation{Simons Society of Fellows, Simons Foundation, New York, NY 10010, USA}
\affiliation{Department of Astrophysical Sciences, Princeton University, Princeton, NJ 08540, USA}
\affiliation{School of Natural Sciences, Institute for Advanced Study, 1 Einstein Drive, Princeton, NJ 08540, USA}
\author{Gerrit S. Farren}
\affiliation{Department of Applied Mathematics and Theoretical Physics, University of Cambridge, Cambridge CB3 0WA, UK}
\author{Blake D. Sherwin}
\affiliation{Department of Applied Mathematics and Theoretical Physics, University of Cambridge, Cambridge CB3 0WA, UK}%
\affiliation{Kavli Institute for Cosmology, Institute of Astronomy, University of Cambridge, Cambridge CB3 0HA, UK}%
\author{Eric J. Baxter}
\affiliation{Institute for Astronomy, University of \hawaii, 2680 Woodlawn Drive, Honolulu, HI 96822, USA}
\author{Dillon J. Brout}\email[NASA Einstein Fellow]{}
\affiliation{Center for Astrophysics | Harvard \& Smithsonian, 60 Garden Street, Cambridge, MA 02138, USA}

%\author{\textit{et al.}}


\begin{abstract} 
Many theoretical resolutions to the so-called ``Hubble tension'' rely on modifying the sound horizon at recombination, $r_s$, and thus the acoustic scale used as a standard ruler in the cosmic microwave background (CMB) and large scale structure (LSS) datasets. As shown in a number of recent works, these observables can also be used to compute $r_s$-independent constraints on $H_0$ by making use of the horizon scale at matter-radiation equality, $k_{\rm eq}$, which has different sensitivity to high redshift physics than $r_s$. \resub{As such, $r_s$- and $k_{\rm eq}$-based measurements of $H_0$ (within a $\Lambda$CDM framework) may differ if there is new physics present pre-recombination.} In this work, we present \resub{the tightest constraints on the latter} from current data, finding $H_0=64.8^{+2.2}_{-2.5}\hun$ at 68\% CL from a combination of BOSS galaxy power spectra, \textit{Planck} CMB lensing, and the newly released \textsc{Pantheon+} supernova constraints, as well as physical priors on the baryon density, neutrino mass, and spectral index. The BOSS and \textit{Planck} measurements have different degeneracy directions, leading to the improved combined constraints, with a bound of $H_0 = 67.1^{+2.5}_{-2.9}$ ($63.6^{+2.9}_{-3.6}$) from BOSS (\textit{Planck}) alone in $\hun$ units. The results show some dependence on the neutrino mass bounds, with the constraint broadening to $H_0 = 68.0^{+2.9}_{-3.2}\hun$ if we instead impose a weak prior on $\sum m_\nu$ from terrestrial experiments rather than assuming $\sum m_\nu<0.26\,\mathrm{eV}$, or shifting to $H_0 = 64.6\pm2.4\hun$ if the neutrino mass is fixed to its minimal value. Even without dependence on the sound horizon, our results are in $\approx 3\sigma$ tension with those obtained from the Cepheid-calibrated distance ladder, \resub{which begins to cause problems for new physics models that vary $H_0$ by changing acoustic physics or the expansion history immediately prior to recombination.}
\end{abstract}

\maketitle

%\section{Introduction}\label{sec: intro}
For better or for worse, the ``Hubble tension" has become one of the key research areas of 
%synonymous with 
twenty-first century cosmology. The problem is straightforward to define: the expansion rate, $H_0$, measured from the local distance ladder calibrated from geometric distances to Cepheid variables stars and type Ia supernovae \citep[e.g.,][]{Riess2022}, is not in agreement with that extracted from the cosmic microwave background (CMB) assuming the standard cosmological model ($\nu\Lambda\mathrm{CDM}$)  \citep{Planck:2018vyg,2020arXiv200710716E}. These measurements depend on different physics: the former relies on the late-time expansion history and certain astrophysical assumptions, whilst the latter is primarily sourced by the baryon acoustic oscillation (BAO) feature in the CMB power spectrum, which depends on the ``sound-horizon'' scale, $r_s$. Despite a wealth of effort, both theoretical and experimental, the problem persists, and, moreover, has become a multidimensional one, with the introduction of a number of new data-sets. These allow alternative probes of the expansion rate, and proceed via a wide number of mechanisms, such as alternative calibration of the distance ladder using tip of the red giant branch (TRGB) methods \resub{\citep[e.g.,][]{2019ApJ...882...34F,2019ApJ...886...61Y,Dhawan:2022yws,Li:2022aho,Freedman:2020dne} or megamasers \citep{2020arXiv201001119B}}, gravitational wave observations \citep[e.g.,][]{2017Natur.551...85A} and time delay cosmography from strongly lensed sources \citep[e.g.,][]{2019MNRAS.498.1420W,2020arXiv200702941B}. For a brief time, $H_0$ constraints appeared to fall in one of two camps: measurements depending on the full $\nu\Lambda$CDM model preferred lower $H_0$, whilst those depending only on local physics tended towards higher values \citep[e.g.,][]{2019NatAs...3..891V}. However, this division has become much less clear (at least in the latter category) with the publication of \resub{new} TRGB and strong lensing results \resub{\citep{2019ApJ...882...34F,2020arXiv200702941B,Dhawan:2022yws,Li:2022aho,Freedman:2020dne}}.

A particularly interesting probe of the expansion rate is that of large scale structure. In the last decade this has generally been analyzed by way of the BAO feature, extracted from the oscillatory part of the galaxy power spectrum \citep[e.g.,][]{Beutler2017,2019JCAP...10..029S,2018MNRAS.480.3879A,deMattia21,2018ApJ...853..119A,2019JCAP...10..044C}. 
%`baryon acoustic oscillation' feature (BAO), resulting from sound waves in the early Universe \citep[e.g.,][]{Beutler2017,2019JCAP...10..029S,2018MNRAS.480.3879A,deMattia21,2018ApJ...853..119A,2019JCAP...10..044C}. 
In combination with external information on the Universe's composition (usually constraints on $\omega_{\rm cdm}$ and $\omega_b$ from the CMB), this feature can be used as a standard ruler to constrain $H_0$, and results in values consistent with those of \textit{Planck} and more recent CMB experiments.  

It has long been known that the matter power spectrum contains information beyond the BAO feature and hence beyond the sound horizon scale. Indeed, a second key physical scale leaves a characteristic imprint on the shape of the matter power spectrum and contains a significant amount of cosmological information: the wavenumber corresponding to the horizon size at matter-radiation equality ($z \approx 3500$), $k_{\rm eq}$. This ``equality scale'' not only sets the scale of the peak of the matter power spectrum but is important for determining the broadband shape of the linear power spectrum at $k\gtrsim k_{\rm eq}$.\footnote{In fact, the equality scale was used as the original source of cosmological information from galaxy surveys \citep[e.g,][]{1997PhRvL..79.3806T,2001MNRAS.327.1297P}).}

Since many models hoping to resolve the ``$H_0$ tension'' proceed by modifying the sound horizon $r_s$ (which sets the BAO scale) \citep[e.g.,][]{DiValentino2021,Knox2020}, extracting $H_0$ constraints by using the equality scale as a standard ruler instead of the sound horizon scale becomes highly desirable. However, it can be challenging to extract Hubble constant information only from $k_{\rm eq}$, because standard analyses of matter power spectrum observables are typically dominated by BAO constraints and hence by sound horizon information.

Several approaches have hence been developed to remove sound horizon information and thus measure the Hubble constant from only the equality scale \resub{(within a $\Lambda$CDM context)}. The first approach was to use the CMB lensing power spectrum \citep{Baxter2020}; since this observable is given by a projection of the matter power spectrum, the BAO oscillations average out such that only equality scale information remains. Later approaches \citep{Farren:2021grl,Philcox2020} improved upon these constraints with novel analyses of the full 3D galaxy power spectrum, building on recent advances in modeling and analyzing the galaxy power spectrum, beyond just its oscillatory component \citep[e.g.,][]{2020JCAP...05..042I,Ivanov2021,Philcox:2021kcw,2020JCAP...05..032P,2020JCAP...05..005D,2021arXiv211007539Z,chen21,Kobayashi:2021oud,Nunes:2022bhn}. Whilst constraints from standard full-shape analyses remain BAO-dominated,  \citep{Philcox2020} successfully removed sound horizon information with a suitable choice of priors (omitting the baryonic information usually provided by Big Bang Nucleosynthesis (BBN) constraints \cite[e.g.,][]{,DES:2017txv,2019JCAP...10..029S,2020JCAP...05..032P,Addison:2017fdm}); subsequently, \citep{Farren:2021grl} proposed and validated a new method to ``integrate out'' the sound horizon even with BBN priors, resulting in the tightest equality scale $H_0$ constraints to date.

The existence of both $k_{\rm eq}$- and $r_s$-derived constraints on $H_0$ allows for interesting consistency checks: in particular, new physics occurring at redshifts of a few thousand will likely cause a discrepancy between the $H_0$ values inferred \resub{by the two $\Lambda$CDM analyses}. As an example, \citep{Farren:2021grl} showed that, \resub{in the context of the Euclid spectroscopic survey}, the early dark energy models preferred by \textit{Planck} \citep{Poulin2019} and ACT data \citep{Hill2021} \resub{(see also \citep{Poulin2021,Moss:2021obd,LaPosta:2021pgm,Smith:2022hwi})}, would induce shifts between the $r_s$- and $k_{\rm eq}$-derived values of $\Delta H_0 = 2.6\hun$ and $7.8\hun$ respectively \resub{(within a $\Lambda$CDM analysis framework)}, \resub{with neither measurement correctly reproducing the input value.}\footnote{\resub{Although these results were obtained from a forecast using Euclid survey parameters, they are roughly independent of the experimental precision, and we expect them to have only weak dependence on the redshift-binning strategies, though they may be influenced by prior volume effects. To assess this, we have repeated the analysis of \citep{Farren:2021grl} for the \textit{Planck} EDE model, finding $\Delta H_0 = 2.4\hun$, when using the BOSS experimental set-up and prior choices.}} So far, the \resub{equality and sound horizon} measurements are in agreement; however, it remains to be seen whether this holds true with the advent of higher precision data.

Motivated by the above, the goal of this work is to place the tightest indirect constraints on the expansion rate from large scale structure observables (galaxy clustering and CMB lensing) within $\nu\Lambda$CDM, but without dependence on the sound horizon. Although previous constraints have been presented in \citep{Farren:2021grl, Philcox2020}, this work extends beyond the former in a number of ways: \resub{(a) we utilize the newest constraints on the matter density from \textsc{Pantheon+} \citep[][]{Brout:2022vxf}, significantly reducing parameter degeneracies, (b) we include marginalization over the sound horizon following \citep{Farren:2021grl}, both for the power spectrum alone and in combination with lensing (unlike \citep{Philcox2020}), (c) we add bounds on the neutrino mass following \textit{Planck} and terrestrial experiments \citep{Planck:2018vyg,Aker:2021gma}, (d) we analyze the latest galaxy power spectra from BOSS, corrected for previous systematic errors, with an updated theoretical model.} As shown below, this yields competitive $r_s$-independent constraints on $H_0$ and leads to an interesting cosmological interpretation. 

\section{Datasets}\label{sec: data}

We begin by discussing the datasets used in this work. Our analysis makes use of four sources of information:
\begin{itemize}
    \item \textbf{CMB Lensing}: We use the \textit{Planck} CMB-marginalized lensing likelihood discussed in \citep{2020A&A...641A...5P,Planck:2018lbu}. This constrains cosmological parameters via the integrated matter density over a broad redshift range from decoupling until today. As discussed in \citep{Baxter2020}, this does not capture information from the sound horizon, due to the smoothing effects of projection integrals. This is implemented in \textsc{montepython} \citep{Brinckmann:2018cvx}, using version R3.10 of the public \textsc{clik} likelihood.
    \item \textbf{Galaxy Power Spectra}: As shown in \citep{Philcox2020,Farren:2021grl}, galaxy power spectra can be used to obtain an $r_s$-independent constraint on $H_0$ either by performing the analysis without a prior on the baryon density or by explicit marginalization over $r_s$ within the likelihood, via a rescaling of the oscillatory component. Here, we adopt the latter strategy. We use the most up-to-date version of the BOSS DR12 galaxy survey dataset \citep{SDSS:2011jap,SDSS-III:2015hof,BOSS:2016wmc,Beutler2017}, using the power spectrum measured in \citep{Philcox:2021kcw} via the window-free estimators of \citep{Philcox:2020vbm,Philcox:2021ukg}. We use BOSS data from two redshift bins (centered at $z = 0.38$ and $z = 0.61$) in two regions of the sky (NGC and SGC), analyzing the unreconstructed power spectrum multipoles (monopole, quadrupole, and hexadecapole) up to $k_{\rm max} = 0.2\Mpch$, as well as the real-space extension, $Q_0$, up to $k_{\rm max} = 0.4\Mpch$ \citep{Ivanov:2021haa}. Data are analyzed using publicly available likelihoods, which implement a theoretical model based on the Effective Field Theory of Large Scale Structure, and marginalize over all necessary nuisance parameters, in addition to the sound horizon.\footnote{Available at \href{https://github.com/oliverphilcox/full_shape_likelihoods}{github.com/oliverphilcox/full\_shape\_likelihoods}.} Note that the galaxy dataset differs slightly from that used in influenced by \citep{Farren:2021grl,Philcox2020}, and corrects a previous error in the normalization. The addition of the large-scale galaxy bispectrum \citep{Philcox:2021kcw,Ivanov:2021kcd} was not found to appreciably improve the parameter constraints.
    \item \textbf{Supernovae (SNe)}: This work constrains $H_0$ by measuring the angular scale of the cosmological horizon at matter-radiation-equality, \textit{i.e.}\ $k_{\rm eq}D_A(z) \propto k_{\rm eq}h$. \resub{Within $\Lambda$CDM,} the equality scale is proportional to $\omega_{cb}\equiv\omega_{cdm}+\omega_b$ \citep{1998ApJ...496..605E}, thus our measurements are necessarily degenerate with the matter density, and can be improved by the addition of $\Omega_m$ priors. Here, we adopt the Gaussian prior $\Omega_m = 0.338\pm0.018$ from the recent \textsc{Pantheon+} analysis \citep{Brout:2022vxf}. Notably, the central value is higher than that of the original \textsc{Pantheon} constraint: $0.298\pm0.022$ \citep{ogpantheon}, which will have implications for the $H_0$ constraints. The origin of this shift in $\Omega_m$ is discussed in \citep[Sec.\,5]{Brout:2022vxf}. These constraints do not fix the supernova absolute magnitude calibration (such as from the local distance ladder), and are thus largely independent of the local distance ladder $H_0$. \resub{We note that, fixing the dark energy of state and ignoring any additional calibration data, the SNe sample primarily measures $\Omega_m$, thus there is little to be gained by utilizing the full \textsc{Pantheon+} posterior rather than just an $\Omega_m$ prior.}
    \item \textbf{Big Bang Nucleosynthesis (BBN)}: To maximize the information that can be extracted from the equality scale, we impose a prior on the physical baryon density $\omega_b = 0.02268 \pm 0.00036$ following \cite{2020JCAP...05..032P}. As shown in \citep{Farren:2021grl}, this does not add sound-horizon-dependence (due to our $r_s$-marginalization), and is additionally not reliant on \textit{Planck}. In the absence of a BBN prior, there is a degeneracy between the equality-based $H_0$ measurements and $\omega_b$ \citep{Farren:2021grl}; however, the dependence of $k_{\rm eq}$ on $\omega_b$ is comparatively shallow, thus for this degeneracy to appreciably affect $H_0$, we would require a large ($\gg 5\sigma$) change in $\omega_b$. This is strongly disfavored given the consistency of BBN and \textit{Planck} $\omega_b$ constraints.%tight constraints from BBN and \textit{Planck}.
    %\eric{maybe worth promoting footnote to main text?  people might worry about tight prior here}\oliver{agreed}
    %\footnote{As shown in \citep{Farren:2021grl}, there is a clear degeneracy between the equality-based $H_0$ measurements and $\omega_b$ in the absence of a BBN prior. However, the dependence of $k_{\rm eq}$ on $\omega_b$ is comparatively shallow, meaning that a large change in $\omega_b$ would be required to appreciably affect $H_0$. This is strongly disfavored given the tight constraints from BBN and \textit{Planck}.}
\end{itemize}

$H_0$ constraints may be tightened by imposing additional physically-motivated parameter constraints, allowing the breaking of important degeneracies. In this work, we utilize the following set of priors (which are carefully chosen so as not to fold in information from the sound horizon):
\begin{itemize}
    \item \textbf{Neutrino Mass}: Equality-based measurements of $H_0$ from CMB lensing are degenerate with the neutrino mass, $\sum m_\nu$ (though those from galaxy analyses are less so). Analysis of the \textit{Planck} dataset excluding lensing information found $\sum m_\nu<0.26\,\mathrm{meV}$ at 95\% CL \citep{Planck:2018vyg}: this motivates the flat prior $m_\nu\in[0,0.26]\,\mathrm{eV}$. We additionally consider a weaker prior $m_\nu\in[0,0.52]\,\mathrm{eV}$ (roughly at the \textit{Planck} $4\sigma$ level), as well as the pessimistic but uninformative choice $m_\nu\in[0,1]\,\mathrm{eV}$. CMB neutrino mass constraints at sub-eV levels primarily arise from lensing effects in the power spectrum, in particular the smoothing of the power spectrum peaks. Since the degree of smoothing of the power spectrum peaks is physically distinct from the peak location, we expect very broad, conservative priors on the neutrino mass (at the several $\sigma$ level) to have only a negligible dependence on the precise value of $r_s$.
    %primarily measured from peak smearing effects in the CMB (rather than peak positions), thus we do not expect this to have significant dependence on the sound horizon scale. This is additionally verified by looking at the \textit{Planck} posteriors: $\sum m_\nu$ is not strongly correlated to $h\omega_{cb}^{-0.25}\omega_b^{-0.125}$, which is a proxy for the sound horizon scale. 
    An alternative choice is to obtain the priors from non-CMB sources, for example terrestrial experiments. Recent results from the \textsc{katrin} analysis give \resub{$\tilde m_\nu^2\lesssim 0.73\,\mathrm{eV}^2$ at 90\% CL in a Bayesian context for the effective electron anti-neutrino mass $\tilde m_\nu$ \citep{Aker:2021gma}}. \resub{Below, we use the \textsc{katrin} posterior as a prior on $\sum m_\nu$, showing this} to yield similar results to the \textit{Planck}-derived priors above. Finally, we will also consider fixing the neutrino mass to its minimal $\nu\Lambda$CDM value of $0.06\,\mathrm{eV}$ (following a number of previous analyses). In all cases, we assume three degenerate massive neutrinos.
    \item \textbf{Spectral Index}: Following the \textit{Planck} lensing-only analyses \citep{Planck:2018lbu}, we adopt a weak Gaussian prior of $n_s = 0.96\pm0.02$ on the primordial spectral slope. This is measured from a comparison of the amplitude of large-scale CMB modes with small-scale ones; since the overall tilt of the spectrum is not expected to be significantly degenerate with the position of the acoustic oscillation features (set by $r_s$), the value of $n_s$ is also not expected to be strongly influenced by the sound horizon.
    Our constraint is also $\approx 3\times$ weaker than that of the \textit{Planck} primary analysis \citep{Planck:2018vyg} and may thus be considered conservative. \resub{Furthermore, it is well motivated by theoretical considerations, since slow-roll inflation requires $n_s$ slightly below unity.} To test the impact of this, we also perform analyses without the $n_s$ prior, and with the $n_s$ prior replaced by a weak $8\%$ prior on the primordial amplitude $A_s$, following \citep{Baxter2020}. Consideration of the \textit{Planck} posterior shows the latter to be $r_s$-independent \citep{Planck:2018vyg}.
    \item \textbf{Sound Horizon Rescaling}: The $r_s$-marginalization procedure discussed in \citep{Farren:2021grl} integrates over a rescaling of the sound horizon using a free parameter $\alpha_{r_s}$. This naturally requires a prior: here, we assume a Gaussian prior of $1.0\pm0.5$ (with $\alpha_{r_s}=1$ giving no rescaling). As can be seen from the previous work \citep{Farren:2021grl}, this is not informative.
\end{itemize}

In this work, parameter constraints are derived by sampling a multivariate likelihood via Markov Chain Monte Carlo (MCMC) methods. This is implemented in \textsc{montepython} \citep{2013JCAP...02..001A,2018arXiv180407261B}, and we sample over the following set of cosmological parameters:
\beq
    \{H_0, \omega_{\rm b}, \omega_{\rm cdm}, \log 10^{10}A_s, n_s, \sum m_\nu\}.
\eeq
We additionally fix the optical depth of reionization, $\tau_{\rm reio}$, to $0.055$, following the \textit{Planck} lensing analyses \citep{Planck:2018lbu}.\footnote{This is of minimal importance since we do not use the CMB primary anisotropies.} To account for various galaxy formation and non-linear effects, the BOSS likelihood also includes a number of nuisance parameters for each subsample; discussion of these can be found in \citep{Philcox:2021kcw} and they are marginalized over in all cases. For each analysis, we run a number of MCMC chains in parallel, assuming them to converge when the Gelman-Rubin diagnostic satisfies $|R-1|<0.05$. \resub{Finally, we note that all analyses are performed using a $\Lambda$CDM theory model, to facilitate robust null tests.}

\section{Results}\label{sec: results}

\begin{figure}
    \centering
    \includegraphics[width=0.7\textwidth]{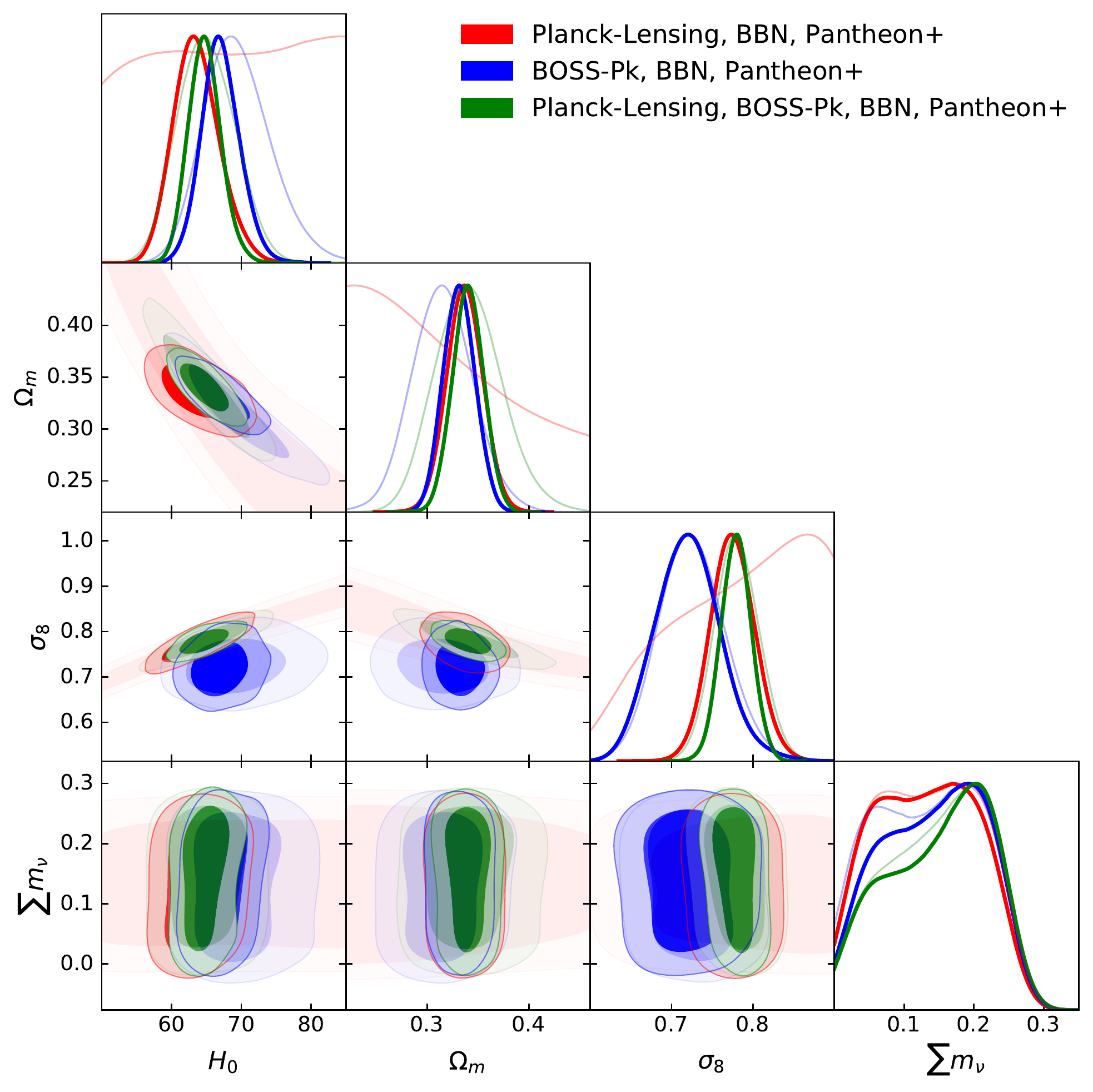}
    \caption{Sound-horizon-independent constraints on the Hubble parameter from CMB lensing (red), galaxy surveys (blue) and their combination (green). All analyses include constraints on $\Omega_m$ from the \textsc{Pantheon+} sample, a prior on $\omega_b$ from BBN, a weak \textit{Planck}-inspired prior on $n_s$, and the neutrino-mass constraint $\sum m_\nu<0.26\,\mathrm{eV}$, matching the \textit{Planck} 95\% limit. The faint lines show the results without a \textsc{Pantheon+} prior on $\Omega_m$; we find clear degradation, particularly for the lensing-only case, which is unable to constrain $H_0$. \resub{Assuming the approximate $k_{\rm eq}-\omega_m/h$ relation from \citep{1998ApJ...496..605E}, the three datasets constrain $k_{\rm eq}=1.60^{+0.07}_{-0.08}$, $1.66\pm0.05$, $1.64\pm0.05$, respectively, in $10^{-2}\hMpc$ units.} Corresponding $H_0$ values are given in Tab.\,\ref{tab:H0vals}. The dark green constraint (combining \textit{Planck}, BOSS and \textsc{Pantheon+}) is the main result of this work.}
    \label{fig: main-plot}
\end{figure}

\begin{table}[]
    \renewcommand{\arraystretch}{1.3}
    \caption{$H_0$ constraints from the analysis of \textit{Planck} lensing and the BOSS DR12 galaxy power spectra, as shown in Fig.\,\ref{fig: main-plot}. Here, we show the results from various data-set combinations, with and without a prior on $\Omega_m$. In all cases, we assume BBN priors on $\omega_b$, weak priors on $n_s$ and a flat prior on the neutrino mass sum with $\sum m_\nu<0.26\,\mathrm{eV}$. The $H_0$ posterior from \textit{Planck} lensing without \textsc{Pantheon+} is unconstraining. The bold entry is the main primary result of this work, and all values are quoted in $\hun$ units at 68\% CL.\label{tab:H0vals}}
    \centering
    \begin{tabular}{l|c|c}
    \hline\hline
     & Fiducial\qquad & Without \textsc{Pantheon+}\\\hline
     \textit{Planck} Lensing & $63.6^{+2.9}_{-3.6}$ & $71^{+20}_{-20}$\\ %\eric{is this just the prior?}\oliver{essentially yes}
     BOSS Galaxies & $67.1^{+2.5}_{-2.9}$ & $69.6^{+4.1}_{-5.4}$\\
     \textit{Planck} Lensing \& BOSS Galaxies\quad\qquad & $\mathbf{64.8^{+2.2}_{-2.5}}$ & $65.0^{+3.9}_{-4.3}$  \\\hline\hline
    \end{tabular}
\end{table}

Fig.\,\ref{fig: main-plot} and Tab.\,\ref{tab:H0vals} show the main results of our analysis: sound-horizon-independent constraints on $H_0$ from CMB lensing and galaxy power spectra, supplemented by priors on $\Omega_m$ from \textsc{Pantheon+} and $\omega_b$ from BBN. In all cases, the upper limit of $\sum m_\nu = 0.26\,\mathrm{eV}$ (the \textit{Planck} $2\sigma$ limit) is assumed. Considering first the CMB lensing results, we find the constraint $H_0 = 63.6^{+2.9}_{-3.6}\hun$ at $68\%$ CL, \resub{or, using the approximate $\Lambda$CDM relation between $k_{\rm eq}$ and $\omega_m/h$ given in \citep{1998ApJ...496..605E}, $k_{\rm eq} = (1.60^{+0.07}_{-0.08})\times 10^{-2}\hMpc$.} If the \textsc{Pantheon+} dataset is removed, the constraining power reduces to almost zero, with the figure showing a strong $\Omega_m-H_0$ degeneracy, as expected from a $k_{\rm eq}$-based measurement. This result is markedly different to that quoted in \citep{Baxter2020}:  $73.5\pm5.3\hun$, both with a lower expansion rate and a significantly tighter errorbar. This arises from two factors: (a) we assume a tighter prior on the neutrino mass that promotes smaller $H_0$ and reduces the posterior width (see below), and (b) the previous study used $\Omega_m$ priors from \textsc{Pantheon}, rather than \textsc{Pantheon+}. As remarked above, the updated supernovae measurements yield a $\approx 1.5\sigma$ higher mean value of $\Omega_m$ and a $20\%$ tighter errorbar (which is more consistent with other recent measurements, such as those of DES Y3 \citep{DES:2021wwk}). Due to the negative $\Omega_m-H_0$ correlation seen in Fig.\,\ref{fig: main-plot}, shifts the result to smaller $H_0$.\footnote{At fixed $k_{\rm eq}$, $\Omega_mH_0$ is constant, thus $\Delta H_0/H_0\approx -\Delta \Omega_m/\Omega_m$. The shift in the central value of $\Omega_m$ moving from \textsc{Pantheon} to \textsc{Pantheon+} is thus expected to induce changes in $H_0$ of approximately $-12\%$, which is consistent with that found in our CMB lensing analyses. When spectroscopic data is included, the shift is expected to reduce, since galaxy power spectra also constrain $\Omega_m$.} Although the posterior is somewhat non-Gaussian, this CMB+SNe measurement is in some tension with the most recent Cepheid $H_0$ results of \citep{Riess2022}.

The $r_s$-marginalized BOSS power spectra constrain $H_0$ to $67.1^{+2.5}_{-2.9}\hun$ \resub{and $k_{\rm eq}=(1.66\pm0.05)\times 10^{-2}\hMpc$} with a factor of $\approx 2$ degradation \resub{in $H_0$} when the \textsc{Pantheon+} prior is removed. Note that LSS power spectra can measure $\Omega_m$ internally (via Alcock-Paczynski distortions, \citep{Ivanov2017}), thus the dataset still retains some constraining power. Previous work found $69.5^{+3.0}_{-3.5}\hun$ \citep{Farren:2021grl}, which is somewhat weaker. In this case the improvements are due to the tighter $\Omega_m$ prior, as well as updates to the new BOSS likelihoods, which include more small-scale data and an improved treatment of the survey window function \citep{Philcox:2021kcw}, \resub{as well as the addition of BBN information relative to \citep{Philcox2020}}. We additionally note that the $\sigma_8$ posterior from BOSS is somewhat below that of the Planck lensing; this could be a manifestation of the claimed ``$S_8$ tension'', \resub{though the individual posteriors remain largely in agreement}. %\eric{seems a bit different from what people usually mean when they refer to $S_8$ tension} \oliver{rewritten slightly}
Finally, we note that these constraints are only around a factor of $(2-3)\times$ weaker than those obtained \textit{including} the sound horizon \citep{Philcox:2021kcw}. This highlights the utility of full-shape information, the importance of which will progressively grow in the next decade \citep{Farren:2021grl}.

The final dataset in Fig.\,\ref{fig: main-plot} is that obtained from the combination of CMB lensing and LSS power spectra. Here, we find a $1\sigma$ constraint of $H_0 = 64.8^{+2.2}_{-2.5}\hun$, \resub{or $k_{\rm eq} = (1.64\pm0.05)\times 10^{-2}\hMpc$}. Na\"ively, one might have expected a $\sqrt{2}$ improvement from combining two independent data-sets with similar $H_0$ posteriors; in practice, this does not occur since both results discussed above use the same set of priors. That said, the inclusion of \textit{Planck} lensing tightens the BOSS posterior by $\approx 15\%$ (or $\approx 30\%$ in terms of survey volume), mostly driven by the somewhat different degeneracy directions of the two in Fig.\,\ref{fig: main-plot}. Although both datasets extract information from the same physical feature, the shape in the $\Omega_m-H_0$ plane is expected to differ \resub{(cf.\,the $\omega_m/h$ contours in Fig.\,\ref{subfig:MCMC_comp})} due to the differences in the angular diameter distance scaling at high- and low-redshifts. We may compare this constraint to that of \citep{Philcox2020}, which finds $70.6^{+3.7}_{-5.0}\hun$. Our results are significantly tighter due to the combination of the reasons above, or in short: better power spectrum modelling including $r_s$-marginalization, more ambitious neutrino mass priors, and new supernova data. The downwards shift in the posterior occurs due to the shift in $\Omega_m$, which also serves to increase the datasets' compatibility, since the lensing analyses now use an $\Omega_m$ posterior consistent with that found from the galaxy power spectrum.

\begin{figure}
    \centering
    \includegraphics[width=0.49\textwidth]{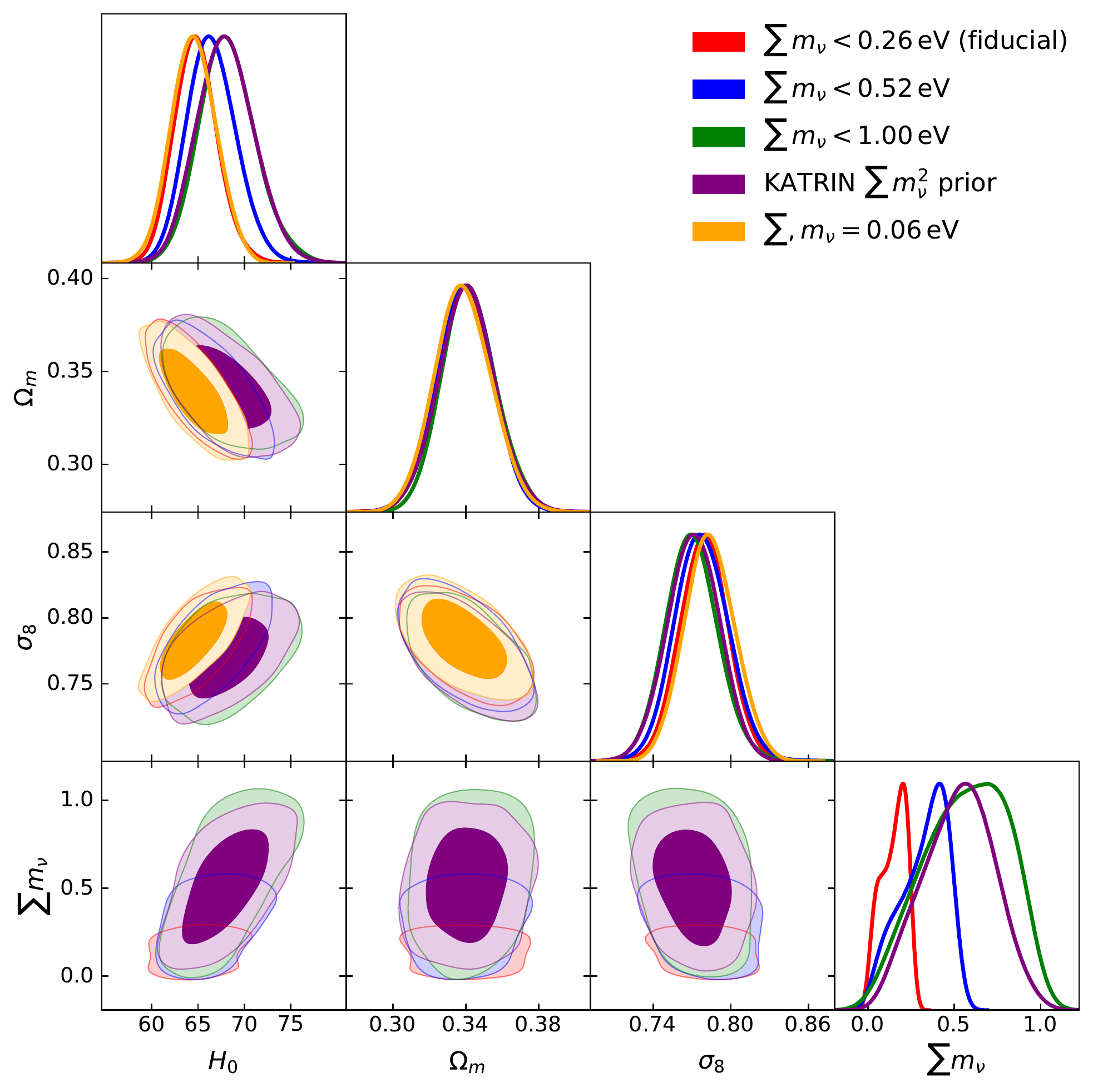}
    \includegraphics[width=0.49\textwidth]{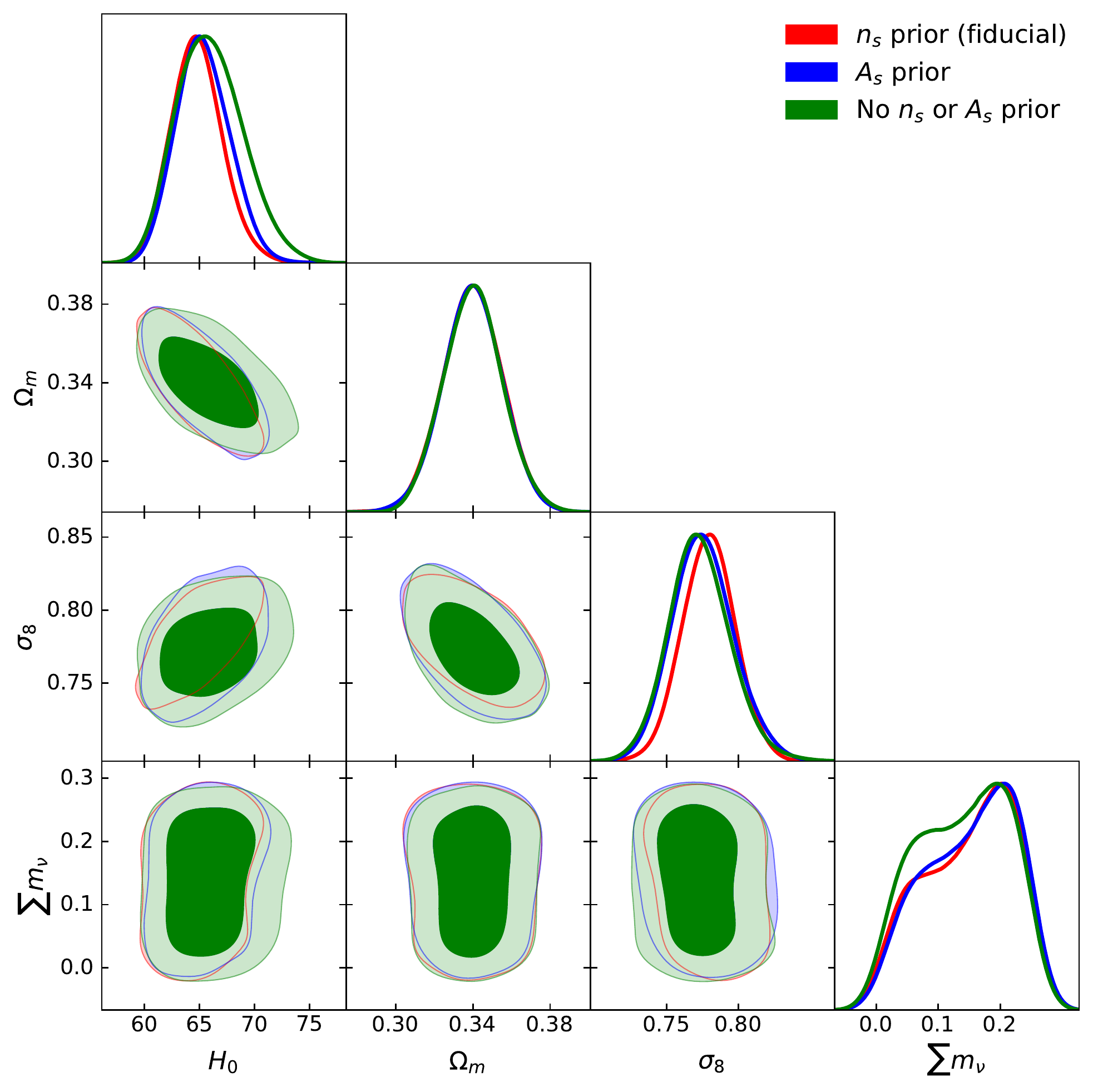}
    \caption{\textbf{Left panel}: As Fig.\,\ref{fig: main-plot}, but examining the dependence on the neutrino mass priors. We show results for an upper bound of $0.26$, $0.52$ and $1.0$\,$\mathrm{eV}$ on the neutrino mass sum in red, blue, and green respectively, and include priors on $\Omega_m$, $\omega_b$ and $n_s$ in all cases. The other contours show analyses including a physical prior on $\sum m_\nu^2$ from ground-based experiments (purple) or fixing the neutrino mass sum to its minimal value, $0.06\,\mathrm{eV}$. Corresponding $H_0$ constraints are given in Tab.\,\ref{tab:H0valsMnu}. \textbf{Right panel}: Dependence of the posterior on the spectral slope prior. We show results from the fiducial analysis (red, with $n_s=0.96\pm0.02$), an analysis with an 8\% prior on $A_s$ (blue), and one with neither an $n_s$ nor $A_s$ prior (green). The three cases have $H_0 = 64.8^{+2.2}_{-2.5}, 65.3^{+2.3}_{-2.6}$ and $66.0^{+2.7}_{-3.4}\hun$ respectively.}
    \label{fig: Mnu-dependence}
\end{figure}

\begin{table}[]
    \renewcommand{\arraystretch}{1.3}
    \caption{$H_0$ constraints from the combination of \textit{Planck} lensing and the BOSS DR12 galaxy power spectra (Fig.\,\ref{fig: Mnu-dependence}) with different choices of neutrino mass prior. We use a flat prior in all cases, except for the penultimate, which is an experimental prior on $\sum m_\nu^2$ from the ground-based \textsc{katrin} experiment, and the last, which fixes the neutrino mass sum to the minimal value of $0.06\,\mathrm{eV}$. All results are given in $\hun$ units.\label{tab:H0valsMnu}}
    \centering
    \begin{tabular}{c|c|c|c|c}
    \hline\hline
    $\sum m_\nu<0.26\,\mathrm{eV}$ & $\sum m_\nu<0.52\,\mathrm{eV}$ & $\sum m_\nu<1.00\,\mathrm{eV}$ & \textsc{katrin}\qquad &$\sum m_\nu = 0.06\,\mathrm{eV}$ \\\hline
    $\mathbf{64.8^{+2.2}_{-2.5}}$ & $66.5^{+2.5}_{-2.8}$  & $68.2^{+2.8}_{-3.2}$ & $68.0^{+2.9}_{-3.2}$ & $64.6\pm2.4$\\\hline\hline
    \end{tabular}
\end{table}

As noted above, our results have significant dependence on the neutrino mass prior. To explore this, we have re-run the joint analysis with two- and four-times wider neutrino mass priors. The corresponding results are displayed in the left panel of Fig.\,\ref{fig: Mnu-dependence} and Tab.\,\ref{tab:H0valsMnu} and demonstrate a clear correlation between $\sum m_\nu$ and $H_0$ \resub{(and a slight preference for $\sum m_\nu>0$ due to prior-volume effects)}. Increasing the width of the flat prior by a factor of four shifts $H_0$ upwards by $3.4\hun$, and increases the width by $30\%$. Some degeneracy is expected, since the effects of neutrino suppression are slightly degenerate with the turnover (and later decline) of the linear power spectrum at $k\gtrsim k_{\rm eq}$, \resub{and further, they degrade any $\omega_m$ information contained within internal $\omega_{cdm}$ constraints}. 
%and $k_{\rm eq}$ and $\sum m_\nu$ influence the high-$k$ power somewhat.% \blake{could also be that both increasing mnu and reducing keq lead to a decrease of high-k power?}. 
%the angular diameter distance (needed to determine the angular scale of $k_{\rm eq}$) depends on $\sum m_\nu$ through  \eric{is this right?  i thought it was because of neutrino suppression being slightly degenerate with turnover in $P(k)$ at $k_{eq}$}\oliver{good point: added this}
As discussed in \citep{Philcox2020}, this effect is significantly stronger for the CMB lensing scenario than for LSS power spectra, and reduces our constraining power. For the reasons described above, however, we consider the $\sum m_\nu<0.26\,\mathrm{eV}$ prior to be conservative within $\nu\Lambda$CDM, and thus regard those measurements as robust. As an additional check, we adopt the CMB-independent prior from \textsc{katrin}. This yields $H_0 = 68.0^{+2.9}_{-3.2}\hun$ from the combination of datasets, with a similar errorbar to the $\sum m_\nu<1\,\mathrm{eV}$ analysis. This result is expected, since the experiment yields an effective $2\sigma$ constraint on the squared mass sum of $0.73\,\mathrm{eV}^2$. To probe the opposite (and less conservative) limit, we consider fixing the neutrino mass sum to its minimal $\nu\Lambda$CDM value; $\sum m_\nu = 0.06\,\mathrm{eV}$. This is a common approximation for analyses with limited dependence on the neutrino mass \citep[e.g.,][]{Farren:2021grl,Planck:2018vyg,Philcox:2021kcw}, and leads to the constraint $H_0 = 64.6\pm2.4\hun$. This is similar to the fiducial ($\sum m_\nu<0.26\,\mathrm{eV}$) case and is primarily dominated by the galaxy survey data, which do not show a strong degeneracy with $\sum m_\nu$ \citep{Philcox2020}.

Finally, we consider the impact of the prior on the spectral slope $n_s$. This is expected to improve the measurements of $k_{\rm eq}$ by fixing the large-scale power spectrum shape; however, it is extracted from a joint fit to the \textit{Planck} primary CMB spectra (but significantly inflated) and thus could, in principle, contain some $r_s$ dependence. Rerunning the joint \textit{Planck}, BOSS and \textsc{Pantheon+} analysis without the $n_s$ prior, we find $H_0 = 66.0^{+2.7}_{-3.4}\hun$, which is $30\%$ broader than the fiducial constraint of $64.8^{+2.2}_{-2.5}\hun$, and a little higher, yet still fully consistent (at $0.6\sigma$, using the approach of \citep{Gratton2020}). As in \citep{Baxter2020}, we may also replace the $n_s$ prior by a prior on the primordial amplitude $A_s$: using a weak $8\%$ prior centered on the \textit{Planck} best-fit (but encompassing both sets of values suggested by the ``$S_8$ tension''), we find $H_0 = 65.3^{+2.3}_{-2.6}\hun$, in good agreement with the fiducial result, as shown in the right panel of Fig.\,\ref{fig: Mnu-dependence}. From these results, we conclude that our measurement of the expansion rate is robust to changes in the $n_s$ prior, since these do not lead to sound-horizon-induced biases in $H_0$. 

\section{$r_s$ independence of $H_0$ constraints}

Previous works have demonstrated that, in the era of Euclid and DESI, analysis of the galaxy power spectra with explicit marginalisation over the sound horizon scale can yield $r_s$-independent constraints on $H_0$ even when information on the baryon density is included via a BBN-derived prior on $\omega_b$ \citep{Farren:2021grl}. The former work also argued that this independence should extend to an analysis of the BOSS dataset with an identical model, \textit{i.e.}\ the scenario considered in this work. Nevertheless, we here demonstrate this explicitly, applying tests devised in \cite{Baxter2020, Philcox2020, Farren:2021grl} to the joint analysis of this work, incorporating the new datasets and prior choices.

\begin{figure}
    \centering
    \subfloat[Full power spectrum]{%
      \includegraphics[width=0.48\textwidth]{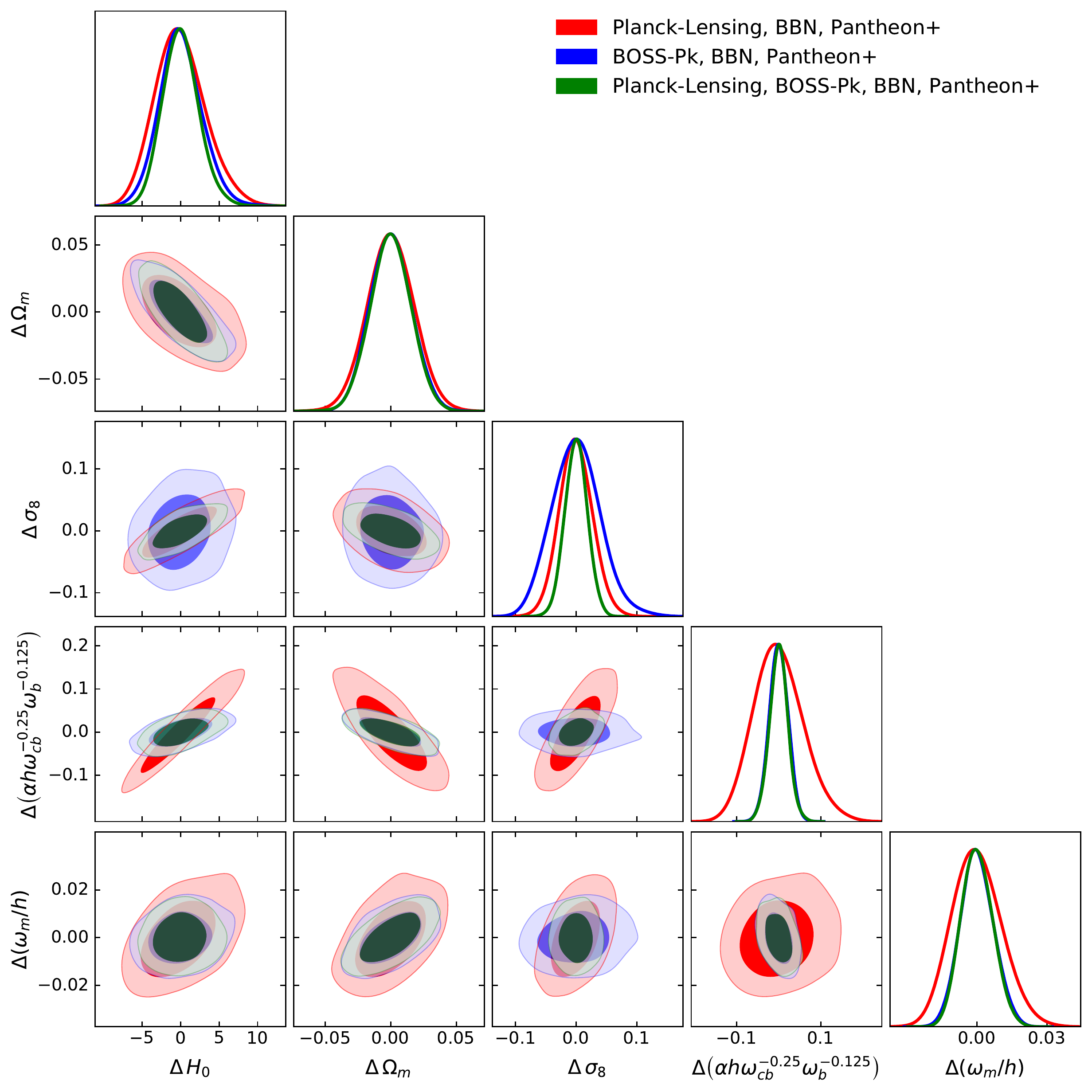}\label{subfig:MCMC_comp}
    }
    \hfill
    \subfloat[Eisenstein-Hu with $r_s$ marginalization]{%
      \includegraphics[width=0.48\textwidth]{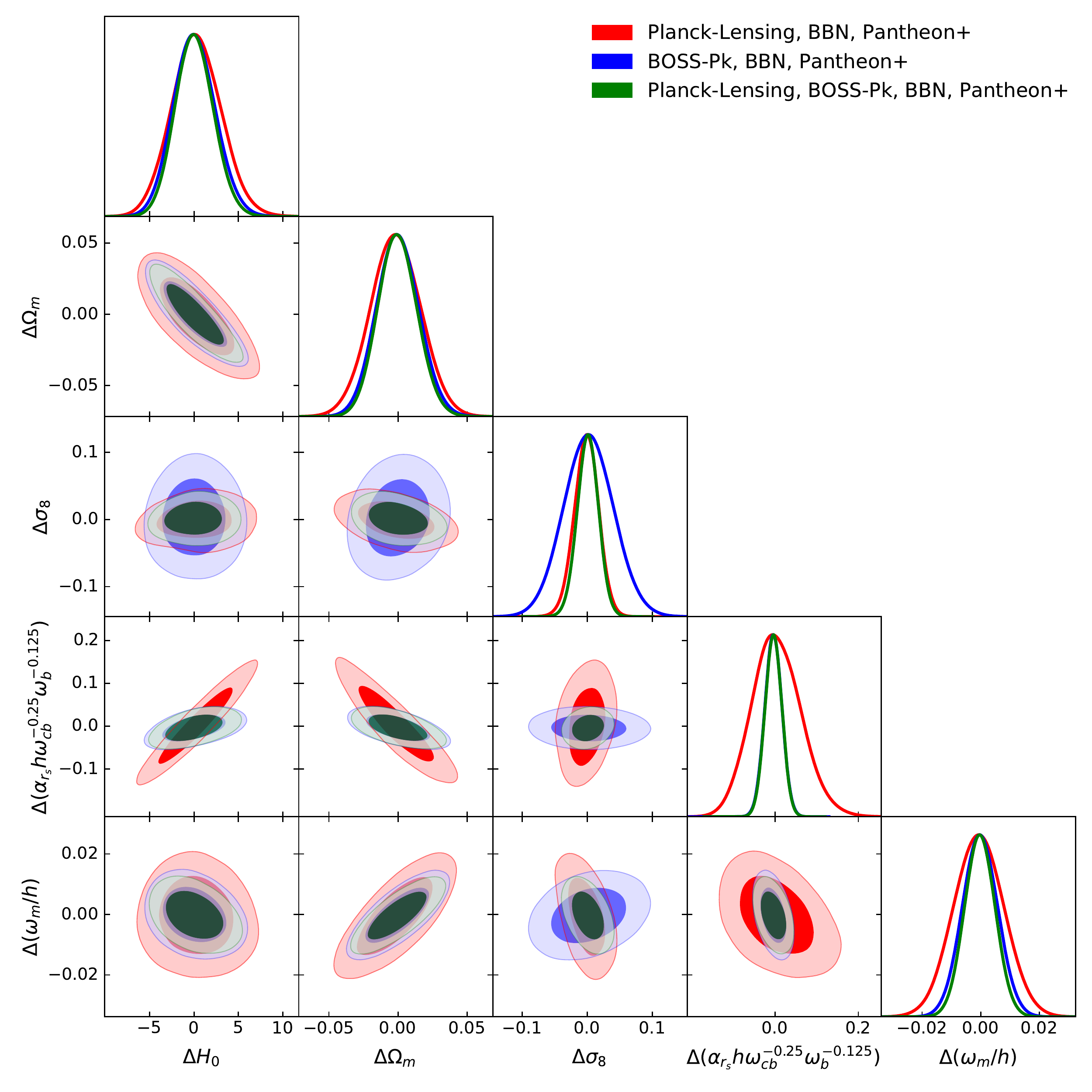}\label{subfig:fisher_comp}
    }
    \caption{Comparison between the fiducial \textit{Planck}, BOSS and joint MCMC analyses (left) to Fisher forecasts run with the same set of parameters and priors, but with explicit marginalization over the sound horizon via an Eisenstein-Hu model (right). In all cases, \textsc{Pantheon+} constraints on $\Omega_m$ are included.
    including a prior on $\omega_b$ from BBN and the %\textsc{Pantheon+} prior on $\Omega_m$ (left) and matching fisher forecasts with an EH model that include exact marginalisation of the sound horizon (right). 
    To simplify the visual comparison, we show only parameter posteriors relative to their mean. We observe an excellent match between the results and forecasts supporting out claim that the sound horizon marginalization procedure is working as expected, and that our constraints are informed primarily by the equality scale. The ``full power spectrum'' results adopt the restrictive prior on the neutrino mass sum ($\sum m_\nu < 0.26 \rm{\ eV}$) while the forecasts neglect neutrinos given that the Eisenstein-Hu model does not include the effect of neutrinos. In addition to $H_0$, $\Omega_m$ and $\sigma_8$ we also show the parameter combinations $\alpha_{r_s} h \omega_{cb}^{-0.25}\omega_{b}^{-0.125}$ ($\alpha_{r_s}$ is the sound horizon rescaling parameter) and $\omega_m/h$; these are roughly proportional to the sound horizon in $\Mpch$-units and the equality scale in $\hMpc$-units respectively.}
    \label{fig:fisher_comparison}
\end{figure}

In Fig.\,\ref{fig:fisher_comparison} we show a comparison between our fiducial MCMC analysis and a Fisher forecast conducted using an Eisenstein-Hu (EH) model for the linear power spectrum \cite{1998ApJ...496..605E}. Within the EH model we are able to explicitly and exactly marginalize over the sound horizon as described in \cite{Farren:2021grl}; this stands in contrast to full Boltzmann computations of the linear power spectrum, in which $r_s$ is an emergent quantity. Notably, we observe a very close match between the full results in Fig.\,\ref{subfig:MCMC_comp} (which include our heuristic $r_s$-marginalization procedure, via a rescaling of the BAO feature) and the EH Fisher forecast with exact marginalization in Fig.\,\ref{subfig:fisher_comp}. To make this comparison, we use the restrictive prior on the neutrino mass ($\sum m_\nu < 0.26 \rm{\ eV}$) for the full analysis since the EH model does not include the effect of neutrinos. Forecast uncertainties on $H_0$ for a range of different dataset combinations with and without the \textsc{Pantheon+} prior on $\Omega_m$ are summarised in Tab.\,\ref{tab:H0_forecasts_EH} which should be compared to Tab.\,\ref{tab:H0vals}.

For the \textit{Planck} lensing forecasts shown in Fig.\,\ref{fig:fisher_comparison}, we do not include explicit marginalization over the sound horizon; initial testing with Fisher forecasts
%\blake{what initial testing? how?} 
found this to have no impact on $H_0$, as predicted in \citep{Baxter2020}. This is additionally consistent with the poor constraints found on the parameter combination $h \omega_{cb}^{-0.25}\omega_{b}^{-0.125}$ (proportional to the sound horizon in $\Mpch$-units) from lensing alone. For the BOSS analysis alone and that joint with \textit{Planck}, this sound-horizon proxy is more well constrained, but we find it to have negligible degeneracy with $H_0$ (see leftmost panel in the second to last row in Fig.\,\ref{subfig:MCMC_comp}), again illustrating $r_s$-independence.

We have also considered the importance of the $r_s$-dependent baryon suppression scale, which we do not marginalize over in either the lensing or spectroscopic analysis. Similarly to \cite{Farren:2021grl}, forecasts show that including this marginalization has only a negligible impact on the Hubble parameter ($\Delta\sigma_{H_0}=0.2\hun$, $\Delta\sigma_{H_0} = 0.1\hun$, and $\Delta\sigma_{H_0} = 0.03\hun$ comparing the forecast uncertainty without any $r_s$ marginalisation and with marginalisation over the power spectrum suppression scale for the lensing-only, galaxy-only and joint analyses respectively). All in all, these checks -- alongside the extensive battery of tests applied to both CMB lensing and galaxy survey probes individually \cite{Baxter2020, Philcox2020, Farren:2021grl} -- support our claim that the information on $H_0$ is not coming from the sound horizon and our analysis thus represents a physically distinct probe of the early Universe.

\begin{table}[]
    \renewcommand{\arraystretch}{1.3}
    \caption{Forecasted uncertainties on $H_0$ from a Fisher forecast matching our joint analysis of \textit{Planck} lensing and the BOSS DR12 galaxy power spectra. These should be compared to the full MCMC results given in Tab.\,\ref{tab:H0vals}, recapitulated below in parentheses. We show the results from various data-set combinations, with and without the \textsc{Pantheon+} constraints on $\Omega_m$. Matching our fiducial analysis, we assume BBN priors on $\omega_b$, and weak priors on $n_s$. The forecast also includes exact marginalization over the sound horizon via an Eisenstein-Hu (EH) model for the BOSS data. We note that we expect our constraints to be marginally tighter than those found in Tab.\,\ref{tab:H0vals} since the EH model does not include the effect of massive neutrinos. All values are quoted in $\hun$ units at 68\% CL.\label{tab:H0_forecasts_EH}}
    \centering
    \begin{tabular}{l|c|c}
    \hline\hline
    & Fiducial & Without \textsc{Pantheon+}\\\hline
     \textit{Planck} Lensing & $\pm 2.8$ ($^{+2.9}_{-3.6}$) & - \\
      BOSS Galaxies & $\pm 2.4$ ($^{+2.5}_{-2.9}$) & $\pm 3.9$ ($^{+4.1}_{-5.4}$)\\
      \textit{Planck} Lensing \& BOSS Galaxies\quad\qquad & $\pm 2.1$ ($^{+2.2}_{-2.5}$)& $\pm 3.2$ ($^{+3.9}_{-4.3}$)   \\\hline\hline
    \end{tabular}
\end{table}

\section{Discussion}\label{sec: conclusion}

The main result of this work is the following \resub{$\Lambda$CDM} constraint on $H_0$ from the equality scale alone: $H_0 = 64.8^{+2.2}_{-2.5}\hun$, using data from BOSS, \textit{Planck} lensing, and \textsc{Pantheon+} supernova constraints, with a weakly restrictive prior on the neutrino mass density. \resub{This does not depend on sound horizon physics, though will be affected by any physical changes to the expansion rate.} In contrast, the most recent Cepheid-calibrated local distance ladder measurement from SH0ES found $73.04\pm1.04\hun$ \citep{Riess2022}, or, when using TRGB calibrators,\footnote{Combining systematic and statistical errors in quadrature.} $69.8\pm1.9\hun$ \citep{2019ApJ...882...34F}, \resub{though there is some disagreement on the TRGB calibration \citep[e.g.,][]{Li:2022aho,Dhawan:2022yws,Freedman:2020dne,Yuan:2019npk}}. Assuming independence, our main result is inconsistent with that of SH0ES at $3.2\sigma$, but agrees with the \resub{CCHP-calibrated} TRGB results at $1.7\sigma$. Furthermore, we find consistency with the main ($r_s$-\resub{dominated}) \textit{Planck} $H_0$ constraints, $H_0 = 67.4\pm0.5\hun$ \citep{Planck:2018vyg} at $1.1\sigma$, \resub{or the $r_s$-dominated full-shape BOSS constraints, $H_0 = 68.3\pm0.8$ \citep{Philcox:2021kcw} at $1.4\sigma$ (though this is not fully independent)}. Strictly, our constraints are not completely independent from those of previous works, since the same supernovae are used in both the distance ladder and our $\Omega_m$ priors. However, the SH0ES $H_0$ posterior shows very weak dependence on $\Omega_m$ \citep[Fig.\,9]{Brout:2022vxf}, implying that any correlation of our results with those of SH0ES is weak.
%dependence \blake{sounds odd... dependence? correlation again?} is weak.
%; furthermore, correlation will in general \textit{reduce} 
%\eric{doesn't this depend on whether it is positive or negative correlation?}\oliver{yes, that's true} 
%the differences between datasets, thus amplifying the above tension. 

The disagreement between our results and those of SH0ES could be resolved in one of two ways: (1) unknown systematics in one or both measurements, or (2), extensions to the standard $\nu\Lambda$CDM model. Importantly, any such extension cannot just alter the sound horizon at recombination, since this would alter the $r_s$-derived $H_0$ constraints but not those of this work. Recent work has largely ruled out the possibility of late-time solutions \citep{DiValentino2021,Knox2020,Dhawan:2020xmp,Benevento:2020fev}, thus a theoretical explanation would likely involve new physics pre-recombination that also affected the equality scale, \textit{i.e.\ }active at redshifts $z\approx 3500$. Furthermore, any new physics \resub{would have to} affect the equality scale and the sound horizon in the same way; unless the modifications occur at very early times (which are themselves constrained by other observations, such as BBN), this seems unlikely, thus our \resub{result disfavors a range} of resolutions \resub{\citep[cf.][]{Farren:2021grl}}, \resub{though is not yet precise enough to strongly constrain models such as the best-fit \textit{Planck} EDE model}. \resub{Of course, new physics that changes the sound-horizon and equality-based constraints in the same manner is not constrained: however, it remains to be seen if such models arise naturally}.
%\blake{Should we emphasize that it would be an odd coincidence (modulo being fitted to Planck) that any new physics would just happen to affect the equality scale and sound horizon in the same way? And that this seems to disfavor new physics resolutions more generally? Up to you how much you want to stress this.}
%For now, systematics seem to be a more likely explanation, though this will be constrained in the future by new and improved $H_0$ measurements, both direct and indirect. 
%\eric{ha...maybe tone down slightly?  or phrase to allow for possibility of systematics in equality/$r_s$ measurements?}.  \oliver{ha okay. I added "indirect"}

Consistency of $H_0$ measurements from the equality and sound-horizon scale provide a useful null test of the $\nu\Lambda$CDM model. 
%\eric{what do we mean by ``our model'' here?  do we mean any new physics model?}.\oliver{Change to the nuLCDM model}
In practice, this can be considered by simply comparing our results to those of the main \textit{Planck} analysis, since the latter is $r_s$-dominated. New physics operating at $z\gtrsim1100$ could lead to a discrepancy in the two scales; 
%\eric{why the upper limit on redshift here?  seems like we could make a slightly more general statement that any new physics model is unlikely to impact the $H_0$ derived from each scale in the same way} \oliver{agreed!}
in \citep{Farren:2021grl} it was shown that this approach could show strong signatures of phenomena such as early dark energy (EDE), with shifts of $\Delta H_0 = 2.6\hun$ ($7.8\hun$) expected from EDE models fit to \textit{Planck} (ACT) data \citep{Smith2020,Hill2021} \resub{(see also \citep{Poulin2021,Moss:2021obd,LaPosta:2021pgm,Smith:2022hwi})} in the context of the Euclid experiment, \resub{or $\Delta H_0=2.4\hun$ for BOSS with a \textit{Planck} EDE model.} In this work, we detect no statistically significant shifts between the $r_s$- and $k_{\rm eq}$-derived datasets using \textit{Planck} and BOSS data. Whilst shifts of a few $\hun$ cannot be ruled out (given the combined errorbar of $\sigma(H_0) = 2.4\hun$), our results place constraints on the more extreme models, and disfavor the best-fit EDE model from ACT. %\blake{while i agree this is very likely, strictly speaking it is not clear that you will get the same delta H0 for BOSS and Euclid analysis and for galaxy only and galaxy + cmb analyses. may want to rephrase a bit more cautiously.} %\eric{how  would this tension show up if you throw in the same datasets we're using + SH0ES into an EDE analysis? I guess you'd get a bad fit?  Is that worth doing?  (maybe not for this paper)}\oliver{I think this goes beyond the scope of this work} 
\resub{A note of caution is in order: to obtain the strongest constraints on a specific new physics realization, one should perform a dedicated analysis with the relevant theoretical model (\citep[e.g.][]{Smith2021,Ivanov2021ede} for EDE). The null tests considered herein allow a broad class of resolutions to be tested simultaneously, albeit at slightly reduced sensitivity.} 

It is interesting to consider how these constraints will develop in the future. From current data, there is little room for improvement; whilst we can push to somewhat smaller scales in the galaxy power spectrum model, these are mostly shot-noise dominated, and do not add significant information about $k_{\rm eq}$. For CMB lensing, the theory is, in general, well understood, thus little improvements can be expected in this case. More promising is the progress expected within the next decade. For the CMB, lensing will be measured to significantly higher precision with AdvancedACT, SPT3G, the Simons Observatory, and CMB-S4 \citep[e.g.,][]{SimonsObservatory:2018koc,Abazajian:2019tiv}, sharpening the lensing-derived $H_0$ contours. This alone will not give particularly large improvements however, with \citep{Baxter2020} forecasting an asymptotic errorbar of $\sigma(H_0) = 3\hun$ from future data when using \textsc{Pantheon+} priors. However, the current methodology may be similarly applied to galaxy lensing, such as with the Rubin observatory. Galaxy surveys will also see a tremendous increase in survey volume, with early data releases from DESI and Euclid expected in the near future. Furthermore, the number of supernovae measured will continue to grow, with forthcoming results expected from surveys such as DES \citep{DES:2022zrg} and the Zwicky Transient Facility \citep{Dhawan:2021hbt}, which will sharpen the $\Omega_m$ prior by a considerable volume. Moreover, neutrino mass constraints will be greatly aided by upcoming CMB and terrestrial experiments, reducing the lensing degeneracies found herein and improving the $H_0$ precision. Whilst it is difficult to forecast exact values on the combination of future data-sets, it is clear that large improvements can be expected; from the Euclid survey alone (without external priors, except from BBN), \citep{Farren:2021grl} predicted $\sigma(H_0) = 0.7\hun$. Coupled with the strong constraints one can place on $H_0$ from the BAO feature, consistency of the two scales will provide a vital cross-check both of our analysis pipelines, and of models of new physics. In this aspect, the future looks exceedingly bright.

\vskip 8pt

\begin{acknowledgments}
\footnotesize
We thank Vivian Poulin, Francis-Yan Cyr-Racine, Niko Šarčević, Anže Slosar for useful feedback, and Eoin Colgain, Florian Niedermann, Martin Sloth and \resub{Tristan Smith} for conversations concerning possible new physics models. \resub{We are additionally grateful to the anonymous referee for providing insightful comments.} OHEP \resub{is a Junior Fellow of the Simons Society of Fellows} and acknowledges funding from the WFIRST program through NNG26PJ30C and NNN12AA01C. GSF acknowledges support through the Isaac Newton Studentship and the Helen Stone Scholarship at the University of Cambridge.
D.B. acknowledges support for this work provided by NASA through NASA Hubble Fellowship grant HST-HF2-51430.001 awarded by the Space Telescope Science Institute (STScI), which is operated by the Association of Universities for Research in Astronomy, Inc., for NASA, under contract NAS5-26555. D.B. thanks the John Templeton Foundation.

The authors are pleased to acknowledge that the work reported in this paper was substantially performed using the Helios cluster at the Institute for Advanced Study. Additional computations were performed using the Princeton Research Computing resources at Princeton University which is a consortium of groups led by the Princeton Institute for Computational Science and Engineering (PICSciE) and the Office of Information Technology's Research Computing Division.

\end{acknowledgments}
\onecolumngrid

\bibliography{bibliography1_Philcox++,bibliography2_Philcox++,bibliography_H0_wo_rd_II}% Produces the bibliography via BibTeX.

\end{document}